\documentclass[runningheads]{llncs}
\usepackage{amsmath}
\usepackage{amssymb}
\usepackage{mathtools}
\usepackage{graphicx}
\usepackage{graphics}
\usepackage{svg}
\usepackage{rotating}
\usepackage{subcaption}

\usepackage{cite}

\usepackage{pifont}
\newcommand{\cmark}{\ding{51}}%
\newcommand{\xmark}{\ding{55}}%

\usepackage[colorinlistoftodos]{todonotes}
\usepackage[colorlinks=true, allcolors=blue]{hyperref}
\usepackage{caption}
\DeclareCaptionFont{mysize}{\fontsize{9}{9}\selectfont}
\usepackage[title]{appendix}
\usepackage{amsfonts}
\usepackage{array,multirow}
\usepackage{mwe}
\usepackage{booktabs}
\usepackage[ruled,vlined,algo2e]{algorithm2e}
\usepackage{xcolor,colortbl}
\definecolor{rred}{rgb}{0.95,0.4,0.3}
\definecolor{bblue}{rgb}{0.4,0.6,1.00}

\usepackage{acro}
\DeclareAcronym{PET}{
short=PET,
long=Positron Emission Tomography
}
\DeclareAcronym{SE}{
short=SE,
long=squeeze and excitation
}

\def\@fnsymbol#1{\ensuremath{\ifcase#1\or *\or \dagger\or \ddagger\or
   \mathsection\or \mathparagraph\or \|\or **\or \dagger\dagger
   \or \ddagger\ddagger \else\@ctrerr\fi}}
\newcommand{\ssymbol}[1]{^{\@fnsymbol{#1}}}

\newcommand\Tstrut{\rule{0pt}{2.4ex}}         
\newcommand\Bstrut{\rule[-0.9ex]{0pt}{0pt}}   

\begin{document}

\title{Synthesizing Multi-Tracer PET Images for Alzheimer's Disease Patients using a 3D Unified Anatomy-aware Cyclic Adversarial Network}

\author{Bo Zhou\inst{1} \and Rui Wang\inst{2,3} \and Ming-Kai Chen\inst{2} \and Adam P. Mecca\inst{4} \and Ryan S. O'Dell\inst{4} \and Christopher H. Van Dyck\inst{4} \and Richard E. Carson\inst{1,2} \and James S. Duncan\inst{1,2} \and Chi Liu\inst{1,2}}
\institute{Department of Biomedical Engineering, Yale University, USA \and Department of Radiology and Biomedical Imaging, Yale University, USA \and Department of Engineering Physics, Tsinghua University, China \and Department of Psychiatry, Yale University, USA}

\authorrunning{B. Zhou, etc} 
\titlerunning{Synthesizing Multi-tracer Brain PET} 

\maketitle              

\begin{abstract}
Positron Emission Tomography (PET) is an important tool for studying Alzheimer's disease (AD). PET scans can be used as diagnostics tools, and to provide molecular characterization of patients with cognitive disorders. However, multiple tracers are needed to measure glucose metabolism ($^{18}$F-FDG), synaptic vesicle protein ($^{11}$C-UCB-J), and $\beta$-amyloid ($^{11}$C-PiB). Administering multiple tracers to patient will lead to high radiation dose and cost. In addition, access to PET scans using new or less-available tracers with sophisticated production methods and short half-life isotopes may be very limited. Thus, it is desirable to develop an efficient multi-tracer PET synthesis model that can generate multi-tracer PET from single-tracer PET. Previous works on medical image synthesis focus on one-to-one fixed domain translations, and cannot simultaneously learn the feature from multi-tracer domains. Given 3 or more tracers, relying on previous methods will also create heavy burden on the number of models to be trained. To tackle these issues, we propose a 3D unified anatomy-aware cyclic adversarial network (UCAN) for translating multi-tracer PET volumes with one unified generative model, where MR with anatomical information is incorporated. Evaluations on a multi-tracer PET dataset demonstrate the feasibility that our UCAN can generate high-quality multi-tracer PET volumes, with NMSE less than $15\%$ for all PET tracers. Our code is available at \href{https://github.com/bbbbbbzhou/UCAN}{https://github.com/bbbbbbzhou/UCAN}.


\end{abstract}
\acresetall

\section{Introduction}
Alzheimer's Disease (AD) is a progressive neurodegenerative disorder and the most common cause of dementia. Positron Emission Tomography (PET), as a functional neuroimaging technique, is commonly used in biomedical research for studying the brain's metabolic, biochemical, and other physiological alterations, thus providing essential information for understanding underlying pathology and helping early diagnosis/differential diagnosis of AD \cite{valotassiou2018spect}. Several important radiotracers have been developed for studying AD's biomarkers, including $^{18}$F-Fluorodeoxyglucose ($^{18}$F-FDG) \cite{cohen2014early}, $^{11}$C-UCB-J \cite{finnema2016imaging}, and $^{11}$C-Pittsburgh Compound-B ($^{11}$C-PiB) \cite{klunk2004imaging}. Specifically, $^{18}$F-FDG provides the metabolic marker that assesses regional cerebral metabolism, and previous studies have revealed that cerebral glucose hypometabolism on $^{18}$F-FDG is a downstream marker of neuronal injury and neurodegeneration \cite{marquez2019neuroimaging}. On the other hand, synapse loss measured by synaptic vesicle protein (SV2A) is another important feature of neurodegeneration \cite{chen2018treatment,nabulsi201411c}. The quantification of SV2A using $^{11}$C-UCB-J serves as another essential marker for synaptic density. Similarly, recent studies have found that the early stage of AD can be characterized by the presence of asymptomatic $\beta$-amyloidosis or increased $\beta$-amyloid burden \cite{sperling2011toward}. Thus, the quantification of $\beta$-amyloid using $^{11}$C-PiB or related tracers is also an indispensable marker for AD studies. Multi-tracer brain PET imaging can provide AD patient with comprehensive brain evaluations on various pathophysiological aspects, enabling more accurate early diagnosis and differential diagnosis as compared to a single tracer study.

However, multi-tracer brain PET is challenging to deploy in real-world scenarios due to the radiation dose to both patient and healthcare providers, as well as the increasing cost when more tracer studies are involved. The ability to synthesize multi-tracer PET images from single-tracer PET images is paramount in terms of providing more information for assessing AD with minimal radiation dose and cost. As $^{11}$C-UCB-J reflects the neural synaptic density, $^{18}$F-FDG reflects the cell metabolism (high activities at the neural synapse) and $^{11}$C-PiB reflecting the neural growth/repair activities, the PET images from these three tracers are correlated which post the possibility to predict one tracer image from another tracer image. With the recent advances in deep learning based image translation techniques \cite{ronneberger2015u,isola2017image}, these techniques have been widely utilized in medical imaging for translation between individual modalities and individual acquisition protocols, such as MR-PET \cite{sikka2018mri}, DR-DE \cite{zhou2018generation}, and PET-PET \cite{wanggeneration}. Even though similar strategies can be utilized for synthesizing multi-tracer PET images from single-tracer PET image, it will result in the need for training large amount of one-to-one domain translation model ($\mathcal{A}_n^2=n \times (n-1)$) and leaving features learned from multiple domain under-utilized. Let $^{18}$F-FDG, $^{11}$C-UCB-J, and $^{11}$C-PiB be denoted as PET tracer A, B, and C, respectively. Given the three PET tracers scenario, traditional methods require $\mathcal{A}_3^2 = 6$ translation models for translating $A \rightarrow B$ / $A \rightarrow C$ / $B \rightarrow A$ / $B \rightarrow C$ / $C \rightarrow A$ / $C \rightarrow B$. In addition, brain PET tracer images, as functional images, are challenging to translate between tracer domain without prior knowledge of anatomy. 

To tackle these limitations, we developed a 3D unified anatomy-aware cyclic adversarial network (UCAN) for translating multi-tracer PET volumes with one unified generative model, where MR with anatomical information is incorporated. Our UCAN is a paired multi-domain 3D translation model based on the previous design of unpaired 2D model of StarGAN \cite{choi2018stargan}. The general pipeline of UCAN is illustrated in Figure \ref{fig:model}. Our UCAN uses training data of multiple tracer domains along with anatomical MR to learn the mapping between all available tracer domains using one single generator. Rather than learning a one-to-one fixed translation, our generator takes tracer volume/MR volume/tracer domain information as input, and learns to flexibly translate the tracer volume into the target tracer domain. Specifically, we use a customized label, a three-channel binary volume with one-hot-channel, for representing tracer domain information. Given trio of tracer volumes, we randomly sample a target tracer domain label and train the generator to translate an input tracer volume into the target tracer domain. During inference stage, we can simply set the tracer domain label and map the original tracer volume into the other tracer domain, hence producing multi-tracer PET volumes from single-tracer PET volume using one generator. Moreover, we proposed to use a 3D Dual Squeeze-and-Excitation Network (DuSE-Net) as our UCAN's generator to better fuse the multiple input information for synthesizing tracer volumes. Extensive experiments on human brain data demonstrate that our UCAN with DuSE-Net can efficiently produce high-quality multi-tracer PET volumes from a single-tracer PET volume using single generator.

\section{Methods}
\begin{figure}[htb!]
\centering
\includegraphics[width=1.00\textwidth]{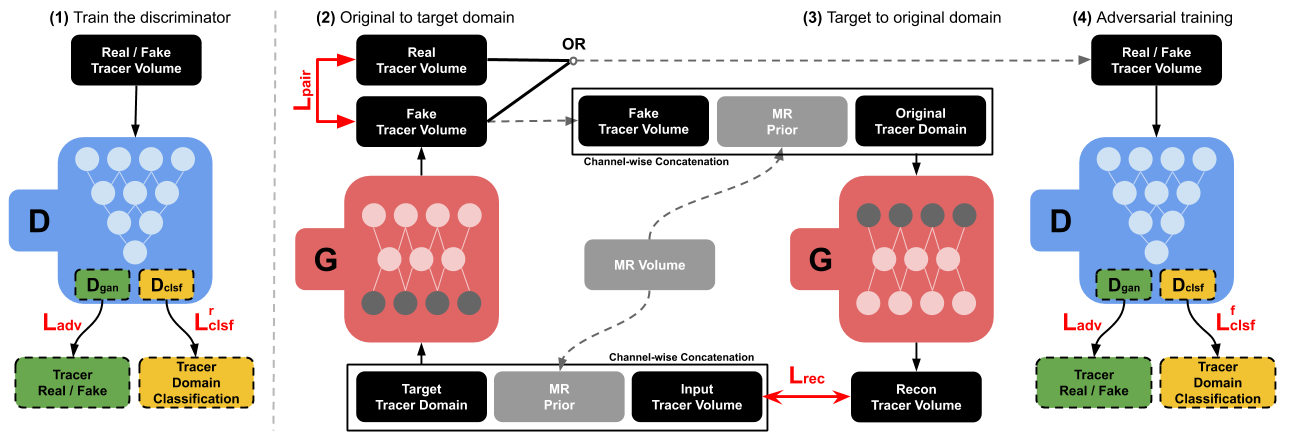}
\caption{Our UCAN consists of one generator $G$ (red block) and one discriminator $D$ (blue block) with detailed structures demonstrated in our supplemental materials. \textbf{(1)} $D$ is trained to discriminate real/fake tracer volumes and classify the real volume's tracer type. \textbf{(2)} $G$ inputs channel-wise concatenation of original tracer volume, MR volume, and target tracer domain label for generation of the target tracer volume. \textbf{(3)} Given predicted target tracer volume, $G$ inputs it along with MR volume and original tracer domain label and tries to reconstruct the original tracer volume. \textbf{(4)} $G$ learns to synthesize fake target tracer volumes that are indistinguishable from real tracer volumes and can be classified into target domain by $D$. The four loss components are marked in \textcolor{red}{\textbf{red}}.}
\label{fig:model}
\end{figure}

The framework and training pipeline of our UCAN is illustrated in Figure \ref{fig:model}. Our UCAN consists of one generator $G$ and one discriminator $D$ for multi-tracer-domain translations. We aim to obtain a single $G$ that can map between tracer domains. Specifically, we train our $G$ to map an input tracer volume $X_{pet}$ into an output tracer volume $Y_{pet}$ conditioned on both the target tracer domain label $M$ and MR volume $X_{mr}$, i.e. $Y_{pet} = G(X_{pet}, X_{mr}, M)$. Specifically, $M$ is a three-channel binary volume with one-hot-channel for target domain labeling, and can be written as
\begin{equation}
    M = [c_1 , c_2, c_3]
\end{equation}
where $[]$ is the channel-wise concatenation operator, and $c_n$ is a one channel binary volume for domain labeling, i.e. $c_1=1, c_2=0, c_3=0$ for tracer A. $M$ is randomly generated over the training process. In parallel, we deploy a discriminator $D$ with auxiliary classifier design \cite{odena2017conditional} for discrimination of source domain (i.e. real/fake) and class domain (i.e. tracer A/B/C). 

\noindent\textbf{Loss Function} To ensure the generated tracer volume lies in target domain and being realistic, we propose four loss components as follows:

\noindent (1) Pair Loss: Given trios of tracer volumes and corresponding MR volume, we randomly sample pairs of tracer volumes along with their MR volume for our pair training. We use a pair loss:
\begin{equation} 
    \mathcal{L}_{pair} = \mathbb{E}[|| G(X_{pet},X_{mr},M) - Y_{pet}^{GT} ||_1] 
\end{equation}
where $G$ generates a tracer volume $G(X_{pet},X_{mr},M) \rightarrow Y_{pet}$ with input of $X_{pet}$ and conditioned on both the MR volume $X_{mr}$ and the target tracer domain label $M$. Here, our $\mathcal{L}_{pair}$ aims to minimize the $L_1$ difference between ground truth target tracer volume $Y_{pet}^{GT}$ and predicted target tracer volume $Y_{pet}$.

\noindent (2) Adversarial Loss: To ensure the synthesized tracer volume is perceptually indistinguishable from the real tracer volume, we utilize an adversarial loss:
\begin{equation} \label{eq:adv}
    \mathcal{L}_{adv} = - \mathbb{E}[log D_{gan}(X_{pet})] - \mathbb{E}[log (1 - D_{gan}(G(X_{pet}, X_{mr}, M)))]
\end{equation}
where $G$ tries to generate realistic tracer volume $G(X_{pet},X_{mr},M) \rightarrow Y_{pet}$ that can fool our discriminator $D_{gan}$, while $D_{gan}$ tries to classify if the input tracer volume is real or fake. That is, $D$ aims to minimize the above equation while $G$ aims to maximize it. In Figure \ref{fig:model}, $D_{gan}$ is one of the distribution output from $D$.

\noindent (3) Classification Loss: While $D_{gan}$ ensures visually plausible tracer volume, given the input volume $X_{pet}$ and a target tracer domain label $M$, the predicted target tracer volume $Y_{pet}$ also needs to be properly classified into the target tracer domain $M$ for synthesizing class-specific tracer appearance. Therefore, we integrate an auxiliary classifier $D_{clsf}$ into our discriminator $D$, as illustrated in Figure \ref{fig:model}. The classification loss can thus be formulated as:
\begin{subequations}
\begin{align}
 \mathcal{L}_{clsf}   & \sim \{ \mathcal{L}_{clsf}^r , \mathcal{L}_{clsf}^f \} \\
 \mathcal{L}_{clsf}^r &= -\mathbb{E}[log D_{clsf}(M^{\dagger}|X_{pet})] \\
 \mathcal{L}_{clsf}^f &= -\mathbb{E}[log D_{clsf}(M|G(X_{pet},X_{mr},M))
\end{align}
\end{subequations}
where $\mathcal{L}_{clsf}$ consists of two parts. The first part of $\mathcal{L}_{clsf}^r$ is the tracer domain classification loss of real tracer volumes for optimizing $D$, which encourages $D$ to learn to classify a real tracer volume $X_{pet}$ into its corresponding original tracer domain $M^{\dagger}$. The second part of $\mathcal{L}_{clsf}^f$ is the loss for the tracer domain classification of fake tracer volumes, which encourages $G$ to generate tracer volumes that can be classified into the target tracer domain $M$.

\noindent (4) Cyclic-Reconstruction Loss: With the adversarial loss and classification loss, $G$ learns to generate tracer volumes that are both realistic and lie in its correct target tracer domain. With the pair loss, $G$ receives direct supervision for synthesizing target tracer volume to learn to preserve content while only alternating the domain-related characteristics. To reinforce the preservation of content over the translation process, we add a cyclic-reconstruction loss:
\begin{equation}
    \mathcal{L}_{rec} =  \mathbb{E}[|| X_{pet} - G(G(X_{pet},X_{mr},M),X_{mr}, M^{\dagger}) ||_1]
\end{equation}
where $G$ inputs the predicted target tracer volume $G(X_{pet}, X_{mr}, M)$, the same MR volume $X_{mr}$, and the original tracer domain label $M^{\dagger}$ for generating the cyclic-reconstructed volume $G(G(X_{pet},X_{mr},M),X_{mr}, M^{\dagger})$ with $G$ utilized twice here. Then, we aim to minimize the $L_1$ difference between original tracer volume $X_{pet}$ and cyclic-reconstructed volume. 

The full objective function consists of four loss components and can be written as:
\begin{equation}
    \mathcal{L}_G = \mathcal{L}_{pair} + \alpha_{rec} \mathcal{L}_{rec} + \alpha_{adv} \mathcal{L}_{adv} + \alpha_{clsf} \mathcal{L}_{clsf}^f
\end{equation}
\begin{equation}
    \mathcal{L}_D = \alpha_{adv} \mathcal{L}_{adv} + \alpha_{clsf} \mathcal{L}_{clsf}^r
\end{equation}
where $\alpha_{clsf}$, $\alpha_{adv}$, and $\alpha_{rec}$ are weighting parameters for classification loss, adversarial loss, and cyclic-reconstruction loss, respectively. In our experiments, we empirically set $\alpha_{clsf}=0.1$, $\alpha_{adv}=0.1$, and $\alpha_{rec}=0.5$. During training, we updated $G$ and $D$ alternatively by: optimizing $D$ using $\mathcal{L}_D$ with $G$ fixed, then optimizing $G$ using $\mathcal{L}_G$ with $D$ fixed. To make the above objective function fully optimized, our $G$ needs to generate realistic target tracer volume with low values on all four-loss components.

\noindent\textbf{Sub-Networks Design} Our discriminator consists of 6 convolutional layers. The first 4 sequential layers extract the features from the input. Then, the feature are simultaneously fed into the last two convolutional layers for classification of real/fake and classification of tracer domain. On the other hand, our generator is a 3D Dual Squeeze-and-Excitation Network (DuSE-Net), and the architecture is based on a 3D U-Net \cite{cciccek20163d} with enhancements on 1.) spatial-wise and channel-wise feature re-calibration \cite{roy2018recalibrating,hu2018squeeze} and 2.) feature transformation in the latent space. Specifically, to better fuse the multi-channel input consisting of tracer domain volume, MR volume, and tracer volume as illustrated in Figure \ref{fig:model}, we propose to use a 3D dual squeeze-and-excitation block at each 3D U-Net's outputs, such that channel-wise and spatial-wise features are iteratively re-calibrated and fused. In the latent space, we also utilize 3 sequential residual convolutional blocks to further transform the feature representation. Network architecture and implementation details are summarized in our supplementary.

\subsection{Evaluation with Human Data}
We collected 35 human studies in our evaluation, including 24 patients diagnosed with AD and 11 patients as healthy control. Each study contains paired brain scans with $^{18}$F-FDG PET scan (tracer A) for glucose metabolism, $^{11}$C-UCB-J PET scan (tracer B) for synaptic density, and $^{11}$C-PiB PET scan (tracer C) for amyloid, and corresponding MR scan for anatomy. All PET scans were acquired on an ECAT HRRT (high-resolution research tomograph) scanner, which is dedicated to PET brain studies, and were performed using $90$-min dynamic acquisitions after tracer injections. Given the difference in kinetics for each tracer, we generated standard uptake value (SUV) images using $60-90$ mins post-injection data for $^{18}$F-FDG, $40-60$ mins post-injection data for $^{11}$C-UCB-J, and $50-70$ mins post-injection data for $^{11}$C-PiB. All PET sinograms were reconstructed into $256 \times 256 \times 207$ volume size with the voxel size of $1.219 \times 1.219 \times 1.231 mm^3$. For each patient, the three tracer PET scans and MR imaging were performed on different days, and the PET images were registered to the MR. The volume size after registration is $176 \times 240 \times 256$ with a voxel size of $1.2 \times 1.055 \times 1.055 mm^3$. We normalized our data by first dividing the maximal value, and then multiplying it by $2$, and then minus $1$ to ensure the intensity lies in $[-1, 1]$. During the evaluation, the data are denormalized. 

We performed five-fold cross validation, where we divided our dataset into a training set of 28 studies and test set of 7 studies in each fold validation. The evaluation was performed on the all 35 studies with all tracer domain translations evaluated. For quantitative evaluation, the performance were assessed using Normalized Mean Square Error (NMSE) and Structural Similarity Index (SSIM) by comparing the predicted tracer volumes and the ground-truth tracer volumes. To evaluate the bias in important brain ROIs, we calculated each ROI's bias using:
\begin{equation}
    Bias = \frac{\sum_{m \in R}Y_{m}^{pred} - \sum_{m \in R}Y_{m}^{gt}}{\sum_{m \in R}Y_{m}^{gt}}
\end{equation}
where $R$ is the specific ROI chosen in the brain. $Y^{pred}$ and $Y^{gt}$ are the predicted tracer volume and ground-truth tracer volume, respectively. For comparative study, we compared our results against the previous translation methods, including cGAN \cite{isola2017image} and StarGAN \cite{choi2018stargan}.

\section{Results}
\begin{figure}[htb!]
\centering
\includegraphics[width=1.00\textwidth]{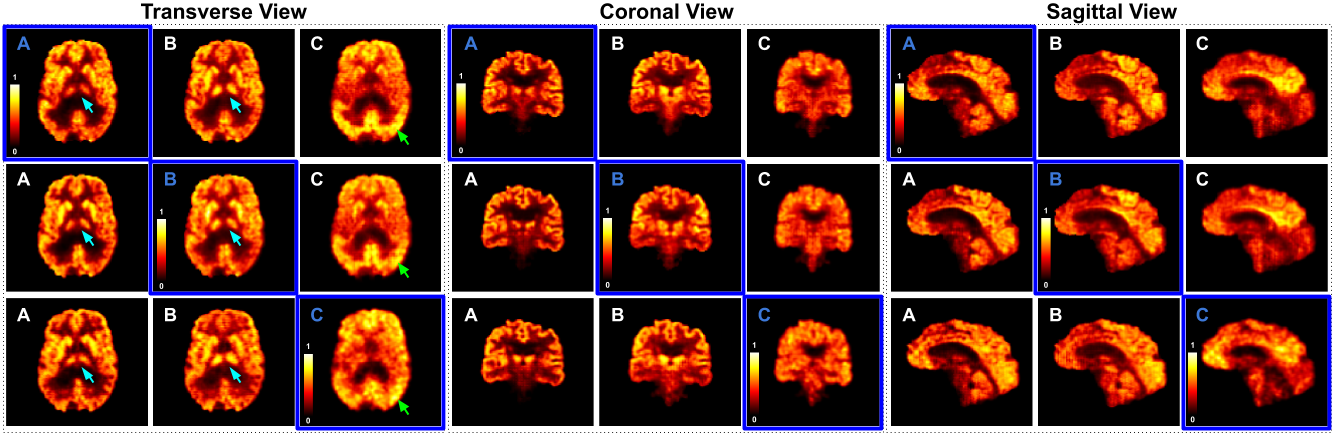}
\caption{Multi-tracer synthesis results using \textbf{UCAN with MR}. Input PET volume (\textcolor{blue}{blue text/box}) and corresponding synthesized PET volumes (white text/box on the same row) are visualization in transverse, coronal, and sagittal 2D views. All volumes are displayed with the same window/level. A: $^{18}$F-FDG, B: $^{11}$C-UCB-J, C: $^{11}$C-PiB.}
\label{fig:vis_1MR}
\end{figure}

The qualitative results of our UCAN is shown in Figure \ref{fig:vis_1MR}. As we can see, given any single tracer volume, our UCAN can generate reconstructions for the remaining two tracers. Because glucose metabolism (measured by tracer A) is positively correlated to the amount of the SV2A (measured by tracer B), the general appearance is similar between image from tracer A and tracer B, except in regions such as thalamus (blue arrows). On the other hand, tracer C ($^{11}$C-PiB) has a low correlation with tracer A ($^{18}$F-FDG) and tracer B ($^{11}$C-UCB-J), the translations are more challenging for $C \rightarrow A$, $C \rightarrow B$, $A \rightarrow C$, and $B \rightarrow C$. While the synthetic results are still not quite consistent with the ground truth globally, we can observe that consistent tracer distribution can be generated in some regions of tracer C (e.g. green arrows). 

Table \ref{tab:comp} outlines the quantitative comparison of different methods. Specifically, we compared our UCAN with the classical one-to-one fixed translation method (cGAN \cite{isola2017image}), and the unified  translation method (StarGAN \cite{choi2018stargan}). Both NMSE and SSIM are evaluated for any two tracers' translations. Our UCAN without MR (UCAN-MR) achieves slightly better performance compared to the conventional one-to-one cGAN, while using only one model. Our UCAN-MR also demonstrates superior performance as compared to the previous unified StarGAN for all the translation paths. Incorporating anatomy information into UCAN significantly improves the translation performance for all the translation paths. For the most challenging translation paths, such as $C \rightarrow A$ and $C \rightarrow B$, we reduced the NMSE from $19.13$ to $10.23$ and $16.81$ to $8.63$, respectively. 

\begin{table} [htb!]
\scriptsize
\centering
\caption{Quantitative comparison of tracer synthesis results using SSIM and NMSE. The optimal and second optimal results are marked in \textcolor{red}{\textbf{red}} and \textcolor{blue}{\textbf{blue}}. A: $^{18}$F-FDG, B: $^{11}$C-UCB-J, C: $^{11}$C-PiB. $\dagger$ means the difference between ours and cGAN are significant at $p<0.05$.}
\label{tab:comp}
    \begin{tabular}{c|c|c|c|c|c|c}
        \hline 
        \textbf{SSIM/NMSE($\times 100 \%$)}         & $A \rightarrow B$ & $A \rightarrow C$ & $B \rightarrow A$ & $B \rightarrow C$ & $C \rightarrow A$ & $C \rightarrow B$ \Tstrut\Bstrut\\
        \hline
        cGAN\cite{isola2017image}$\times 6$         & .767/16.91        & .714/19.13        & .797/19.21        & .689/23.77        & .752/24.26        & .779/18.81        \Tstrut\Bstrut\\
        \hline
        StarGAN\cite{choi2018stargan}               & .701/36.83        & .585/40.13        & .601/43.23        & .574/38.99        & .609/45.20        & .607/37.16        \Tstrut\Bstrut\\
        \hline
        \hline
        UCAN-MR                                     & \textcolor{blue}{.789/14.76}        & \textcolor{blue}{.728/18.82}        & \textcolor{blue}{.814/15.81}        & \textcolor{blue}{.691/19.98}        & \textcolor{blue}{.823/19.13}        & \textcolor{blue}{.802/16.81}        \Tstrut\Bstrut\\
        \hline
        UCAN+MR                                     & \textcolor{red}{.841/8.94}$^\dagger$         & \textcolor{red}{.764/14.58}$^\dagger$        & \textcolor{red}{.899/8.21}$^\dagger$         & \textcolor{red}{.773/13.64}$^\dagger$        & \textcolor{red}{.871/10.23}$^\dagger$        & \textcolor{red}{.821/8.63}$^\dagger$         \Tstrut\Bstrut\\
        \hline
    \end{tabular}
\end{table}

\begin{table} [htb!]
\scriptsize
\centering
\caption{Ablation study on our UCAN with different configurations. \textcolor{green}{\cmark} and \textcolor{red}{\xmark} means module used and not used in our UCAN.}
\label{tab:abla}
    \begin{tabular}{c|c|c|c|c|c|c|c|c}
        \hline 
        \textbf{SSIM/NMSE}                      & $\mathcal{L}_{pair}$      & DuSE                      & $A \rightarrow B$ & $A \rightarrow C$ & $B \rightarrow A$ & $B \rightarrow C$ & $C \rightarrow A$ & $C \rightarrow B$  \Tstrut\Bstrut\\
        \hline
        \multirow{4}{*}{UCAN-MR}                & \textcolor{red}{\xmark}   & \textcolor{red}{\xmark}   & .710/34.41        & .589/39.36        & .608/41.64        & .592/38.68        & .621/43.41        & .612/35.00         \Tstrut\Bstrut\\
                                                & \textcolor{green}{\cmark} & \textcolor{red}{\xmark}   & .779/16.99        & .718/25.63        & .811/18.32        & .620/25.12        & .804/26.24        & .771/21.20         \Tstrut\Bstrut\\
                                                & \textcolor{red}{\xmark}   & \textcolor{green}{\cmark} & .774/17.59        & .611/39.19        & .810/20.85        & .600/38.11        & .647/36.55        & .747/28.01         \Tstrut\Bstrut\\
                                                & \textcolor{green}{\cmark} & \textcolor{green}{\cmark} & .789/14.76        & .728/18.82        & .814/15.81        & .691/19.98        & .823/19.13        & .802/16.81         \Tstrut\Bstrut\\
        \hline
        \multirow{4}{*}{UCAN+MR}                & \textcolor{red}{\xmark}   & \textcolor{red}{\xmark}   & .766/22.12        & .647/24.61        & .807/23.98        & .640/22.56        & .821/21.57        & .800/17.91         \Tstrut\Bstrut\\
                                                & \textcolor{green}{\cmark} & \textcolor{red}{\xmark}   & .798/10.44        & .724/18.84        & .866/9.03         & .742/17.63        & .847/12.96        & .808/12.89         \Tstrut\Bstrut\\
                                                & \textcolor{red}{\xmark}   & \textcolor{green}{\cmark} & .778/16.91        & .653/23.67        & .811/19.31        & .649/21.57        & .828/20.02        & .802/16.97         \Tstrut\Bstrut\\
                                                & \textcolor{green}{\cmark} & \textcolor{green}{\cmark} & .841/8.94         & .764/14.58        & .899/8.21         & .773/13.64        & .871/10.23        & .821/8.63          \Tstrut\Bstrut\\
        \hline
    \end{tabular}
\end{table}

We performed ablation studies on our UCAN with different configurations, including with or without pair loss, 3D DuSE block, and MR information. The quantitative results are summarized in Table \ref{tab:abla}. UCAN without pair loss is difficult to generate the correct tracer image with low correlation, such as $C \rightarrow A/B$. Adding 3D DuSE block along with pair loss allows UCAN to better fuse the input information, thus generating better synthesis results. UCAN with pair loss, DuSE block, and MR information provides the best synthesis results. The corresponding visualization comparison are included in our supplementary.

\begin{figure}[htb!]
\centering
\includegraphics[width=0.96\textwidth]{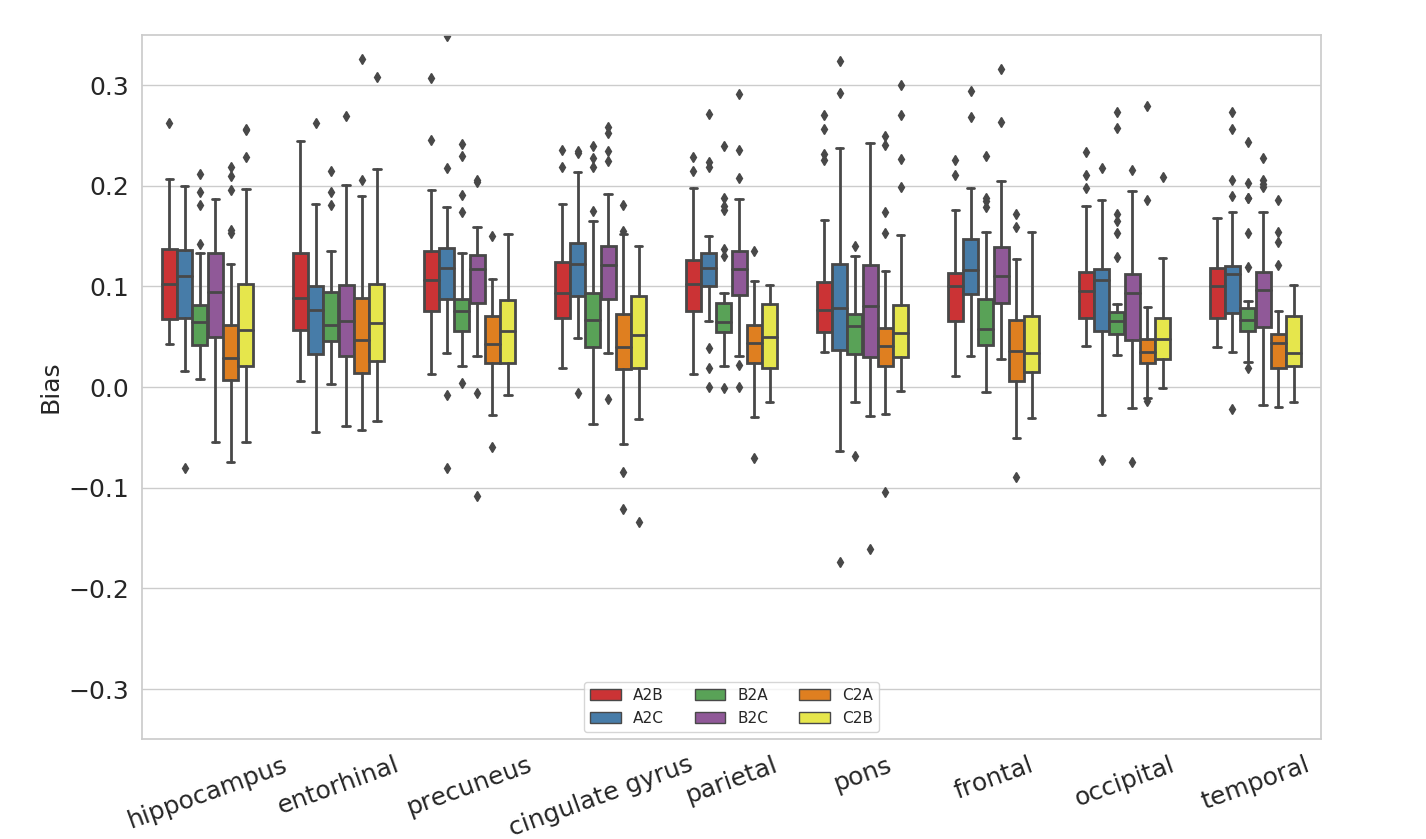}
\caption{Quantitative evaluation of bias in 9 ROIs for UCAN+MR. Mean and standard deviation of bias in 9 AD-related brain ROIs is illustrated for the synthesis between all three tracers. A: $^{18}$F-FDG, B: $^{11}$C-UCB-J, C: $^{11}$C-PiB.}
\label{fig:eval}
\end{figure}

In parallel, we evaluated the bias of our UCAN+MR results in 9 important ROIs, and the results are summarized in Figure \ref{fig:eval}. From the results, the mean bias and associated standard deviation are no more than 15\% for all the ROIs, including hippocampus and entorhinal cortex which are two of the first areas impaired by AD and is an important region for AD early diagnosis. 

\section{Conclusion}
In this work, we proposed a 3D unified anatomy-aware cyclic adversarial network (UCAN), a framework for translating multi-tracer PET volumes with one unified generative model. Our UCAN consisting of four key loss components and domain label volume allows us to take tracer volume/MR volume/tracer domain label as input, and learns to flexibly translate the tracer volume into the target tracer domain. Moreover, the DuSE-Net in our UCAN allows us to better fuse multiple input information for the unified synthesis tasks. Preliminary evaluation using human studies suggested the feasibility that our method is able to generate high-quality multi-tracer PET volumes with acceptable bias in important AD-related ROIs.

\bibliographystyle{splncs}
\bibliography{bibliography}
\end{document}